\documentstyle[prd,aps,eqsecnum,epsf]{revtex}

\makeatletter
\def\figurenum#1{\def\thefigure{#1}\let\@currentlabel\c@thefigure
\addtocounter{figure}{\m@ne}}
\def\fnum@figure{{\rm Fig.\space\thefigure.}}
\makeatother

\newcommand{\bm}{\boldmath}

\begin{document}
\setlength{\baselineskip}{6mm}

\title{Relativistic Zel'dovich approximation
in spherically symmetric model}

\author{Masaaki Morita $^1$, Kouji Nakamura $^2$ and Masumi Kasai $^3$}

\address{
   $^1$ Department of Physics, Tokyo Institute of Technology,
        Oh-Okayama, Meguro-ku, \\ Tokyo 152, Japan \\
   $^2$ Advanced Science Research Center,
        Japan Atomic Energy Research Institute, \\
        Tokai, Naka, Ibaraki 319-11, Japan \\
   $^3$ Department of Physics, Hirosaki University, 
        Bunkyo-cho, Hirosaki 036, Japan
}

\date{\today}

\maketitle

\begin{abstract}

We compare relativistic approximation methods, which describe
gravitational instability in the expanding universe, 
in a spherically symmetric model.
Linear perturbation theory, second-order perturbation theory, 
relativistic Zel'dovich approximation,
and relativistic post-Zel'dovich approximation are considered
and compared with the Lema\^{\i}tre-Tolman-Bondi solution
in order to examine the accuracy of these approximations.
We consider some cases of inhomogeneous matter distribution
while the homogeneous top-hat model has been usually taken
in the previous Newtonian works.
It is found that the Zel'dovich-type approximations are generally
more accurate than the conventional perturbation theories
in the weakly nonlinear regime.
Applicable range of the Zel'dovich-type approximations
is also discussed.

\end{abstract}


%%%%%%%%%%%%%%%%%%%%%%%%%%%%%%%%%%%%%%%
\section{Introduction}
%%%%%%%%%%%%%%%%%%%%%%%%%%%%%%%%%%%%%%%
Structure formation in the universe is an important subject 
of research in cosmology.
A standard view of the structure formation 
is that density fluctuations with small amplitudes 
in the early universe have grown to be a variety of 
cosmic structures due to gravitational instability.
The growth of the density fluctuations has been throughly
investigated by linear perturbation theory of
the Friedmann-Lema\^{\i}tre-Robertson-Walker (FLRW) universe
within both the Newtonian theory and general relativity~\cite{peebles80}.
Relativistic linear perturbation theory was first derived
by Lifshitz~\cite{lifshitz}.
Such relativistic treatments are indispensable when we consider
large-scale fluctuations.
In his theory, however, there remains a gauge problem
that unphysical perturbations are included in the solutions.
This problem was carefully studied later
by Press and Vishniac~\cite{press}.
Also developed was the gauge-invariant formulation~\cite{bardeen},
which gives a conceptually straightforward way
for dealing with cosmological perturbation.

It is true that the linear theories play an important role
in the study of gravitational instability,
but they are valid only in the region where density contrast
$\delta \equiv (\rho - \rho_b)/\rho_b$ is much smaller than unity.
($\rho$ is energy density of the perturbed FLRW universe and
$\rho_b$ is that of the background FLRW universe.)
As $\delta$ grows to be comparable to unity,
nonlinear effects become essential and
we need some kinds of nonlinear approximations.
Tomita~\cite{tomita67} developed second-order perturbation theory
by extending Lifshitz's work to study nonlinear effect
of gravitational instability for the matter-dominated universe.
His approach, however, still depends on the assumption
of $\delta$ being small.
An approximation scheme without the assumption
was proposed by Zel'dovich~\cite{zel70} within the Newtonian framework.
This scheme is known as Zel'dovich approximation,
which is now widely applied to the problems of the large-scale
structure formation.
It has been shown that the Zel'dovich approximation can be regarded
as a subclass of the first-order solutions in the Lagrangian perturbation
theory~\cite{buchert92}.
Then higher-order extension of the Zel'dovich approximation, say,
post-Zel'dovich approximation (and post-post-Zel'dovich approximation
and so on), is straightforwardly derived via the higher-order Lagrangian
approach~\cite{bueh,buchert94,bouchet}.
Relativistic versions of the Zel'dovich approximation have been
also studied for the last few years
by several authors~\cite{cpss,kasai95,rmkb,mate}.
Here we will focus on our tetrad-based approach,
whose correspondence to the original Zel'dovich approximation
is made clear in Ref.~\cite{kasai95} and extension to second order
is presented in Ref.~\cite{rmkb}.

One of remarkably advantageous points of the Zel'dovich-type approximations,
both the original Newtonian one and the relativistic version,
is that they include exact solutions when the deviation from
the background FLRW universe is locally one-dimensional.
These exact solutions are known as Zel'dovich solutions~\cite{zel70}
in the Newtonian case and (some class of)
Szekeres solutions~\cite{szekeres,krasin}
in the general relativistic case, respectively.
For this reason, the Zel'dovich-type approximations are presumably
accurate in description of nearly one-dimensional collapse.
It is not clear, however, whether they also give high accuracy
in the case of non one-dimensional collapse.
In the Newtonian framework, it has been investigated
by using spherical models: The so-called top-hat collapse
model~\cite{munshi}, the top-hat void model~\cite{sasha},
and some more general case~\cite{bouchet}.
(See also Ref.~\cite{saco} for review.)
In addition, there is also a recent work~\cite{ayako}
in which homogeneous spheroidal models are considered.
An interesting implication is obtained in it:
As the deviation of the models from the spherical symmetry becomes larger,
the accuracy of the Zel'dovich-type approximations increases
while the conventional (Eulerian) approximations have the opposite tendency.
It indicates that the Zel'dovich-type approximations may be the least
accurate in the exactly spherical case.
Then, considering the spherical case may tell us the lowest accuracy
of the Zel'dovich-type approximations.

In general relativity, an exact solution of the spherically symmetric
dust model is known as
the Lema\^{\i}tre-Tolman-Bondi (LTB) solution~\cite{landau}.
It is, therefore, of interest to test the Zel'dovich-type approximations
with the exact solutions to examine accuracy of the approximations.
It is also instructive to clarify the differences
between the conventional perturbation theories and the Zel'dovich-type
approximations by imposing spherical symmetry and comparing them
with the exact solutions.

In this paper, we compare the relativistic approximations
such as the linear perturbation theory by Lifshitz~\cite{lifshitz},
the second-order perturbation theory by Tomita~\cite{tomita67},
the relativistic Zel'dovich approximation by Kasai~\cite{kasai95},
and the relativistic post-Zel'dovich approximation
by Russ {\it et al.}~\cite{rmkb} with the LTB solution.
We consider some initial conditions which have inhomogeneous
matter distribution, not homogeneous one like the top-hat model
adopted in the Newtonian works.
It will be shown that the Zel'dovich-type approximations are
more useful than the conventional ones in the quasi-nonlinear regime
in each case.

The plan of this paper is as follows.
In the next section,
we summarize the relativistic perturbation theories mentioned above.
In the section~\ref{spher},
the LTB solution is introduced and the relation
to the relativistic perturbation theories is considered.
Main results of this paper are shown in the section~\ref{comparison}.
The section~\ref{summary} contains summary of our results
and discussions.

Throughout this paper, units are chosen so that $c=1$.
Indices $\mu, \nu, \cdots$ run from $0$ to $3$ and
$i, j, \cdots$ run from $1$ to $3$.

%%%%%%%%%%%%%%%%%%%%%%%%%%%%%%%%%%%%%%%%%%%%%%%%%%%%%%%%%%%
\section{Relativistic perturbation theories}\label{perturb}
%%%%%%%%%%%%%%%%%%%%%%%%%%%%%%%%%%%%%%%%%%%%%%%%%%%%%%%%%%%
In this section, we summarize general relativistic perturbation
theories which describe gravitational instability
in the matter-dominated FLRW universe.
We assume that the background is the Einstein-de Sitter spacetime,
whose line element is
\begin{equation}\label{FLRW}
ds^2 = -dt^2 + a^2(t)\,(dR^2 + R^2 d\theta^2 + R^2 \sin^2\theta d\phi^2)
\equiv -dt^2 + a^2(t)\,k_{ij}\,dx^i dx^j \;,
\end{equation}
where $a(t)=t^{2/3}$ is the scale factor.
The background four-velocity and energy density of the matter are
$u_b^{\mu}=(1,0,0,0)$ and $\rho_b = 1/(6\pi G\, t^2)$, respectively.
Density contrast $\delta$ shown later is defined as
$\delta \equiv (\rho - \rho_b)/\rho_b$,
where $\rho$ is energy density of the perturbed FLRW universe.

%%%%%%%%%%%%%%%%%%%%%%%%%%%%%%%%%%%%%%%%%%%%%%%%%%%%%%%%%%
\subsection{Conventional linear and second-order theories}
%%%%%%%%%%%%%%%%%%%%%%%%%%%%%%%%%%%%%%%%%%%%%%%%%%%%%%%%%%
Lifshitz~\cite{lifshitz} pioneered linear perturbation
of the FLRW universe in the synchronous gauge
\begin{equation}
ds^2 = -dt^2 + g_{ij} \,dx^i dx^j  \;.
\end{equation}
In his theory, the scalar-mode solutions for pressureless
matter (dust) are
\begin{equation}\label{lif}
\left\{ \begin{array}{l}
\gamma_{ij} \equiv a^{-2} g_{ij}
 = \left(1+\frac{20}{9} \Psi \right) k_{ij}
 + 2 t^{\frac23} \Psi_{|ij} + 2 t^{-1} \Phi_{|ij} \;, \\ \\
u^i_{(1)} = 0 \;, \\ \\
\delta_{(1)} = -t^{\frac23} \Psi^{|k}_{\ |k} 
               -t^{-1} \Phi^{|k}_{\ |k} \;,
\end{array}\right.
\end{equation}
where $\Psi=\Psi(\mbox{\bm $x$})$ and $\Phi=\Phi(\mbox{\bm $x$})$
are spatial arbitrary functions of first-order smallness,
$|$ denotes the covariant derivative associated with
the background three-metric $k_{ij}$,
and $u^i$ represents spatial component of the four-velocity of the matter.
Subscript $(1)$ denotes first-order perturbation quantity.
We will not consider contribution of the decaying mode,
which is proportional to $t^{-1}$, later on.

Tomita~\cite{tomita67} developed second-order perturbation theory
by extending Lifshitz's work.
He obtained the following second-order perturbative solutions
from the first-order scalar mode solutions:
\begin{equation}\label{tomi}
\left\{ \begin{array}{l}
\gamma_{ij} = 
\left(1 + \frac{20}{9}\Psi + \frac{100}{81}\Psi^2 \right)k_{ij} 
 + t^{\frac23} \left(2\Psi_{|ij} - \frac{40}{9}\Psi_{|i}\Psi_{|j}
 -\frac{20}{9}\Psi\Psi_{|ij}+\frac{10}{9}\Psi^{|k}\Psi_{|k}\,k_{ij} 
          \right) \\
\hspace*{10mm}
    + \frac17 t^{\frac43} \left[ 19 \Psi_{|ik}\Psi^{|k}_{\ |j} 
    - 12 \Psi^{|k}_{\ |k}\Psi_{|ij} 
    + 3 \left( \left(\Psi^{|k}_{\ |k}\right)^2 
    - \Psi^{|k}_{\ |\ell}\Psi^{|\ell}_{\ |k} \right) k_{ij} 
                         \right] \;, \\ \\
u^i_{(1)} = 0 \;, \quad u^i_{(2)} = 0 \;, \\ \\
\delta_{(1)} + \delta_{(2)} = -t^{\frac23} \Psi^{|k}_{\ |k}
    +\frac59 t^{\frac23} \left(
     \Psi^{|k} \Psi_{|k} + 6\Psi \Psi^{|k}_{\ |k}
                         \right)
    +\frac17 t^{\frac43} \left(
     5(\Psi^{|k}_{\ |k})^2 + 2\Psi^{|k}_{\ |\ell}\Psi^{|\ell}_{\ |k}
                         \right) \;.
\end{array}\right.
\end{equation}
Subscript $(2)$ represents second-order perturbation.
Here we neglected the second-order tensor mode, which is induced
by the first-order scalar mode and does not appear in the spherical case.

%%%%%%%%%%%%%%%%%%%%%%%%%%%%%%%%%%%%%%%%%%%%%%%%%%%%%%%%%%%%%%%%%
\subsection{Zel'dovich-type approximations in general relativity}
%%%%%%%%%%%%%%%%%%%%%%%%%%%%%%%%%%%%%%%%%%%%%%%%%%%%%%%%%%%%%%%%%
In this subsection, we review a relativistic version of the
Zel'dovich approximation developed by us~\cite{kasai95,rmkb}.
The irrotational dust model is assumed and then
we can take the comoving synchronous coordinate
\begin{equation}
ds^2 = -dt^2 + g_{ij} \,dx^i dx^j
\end{equation}
with the four-velocity $u^{\mu}=(1,0,0,0)$.
Thanks to this choice of the gauge, the energy equation
$u_{\mu}T^{\mu\nu}_{\ \ ;\nu}=0$ with
$T^{\mu\nu} = \mbox{diag}[\rho,0,0,0]$, which becomes
\begin{equation}
\dot\rho + \rho K^i_{\ i} = 0 \;,
\end{equation}
is formally solved in the form
\begin{equation}
\label{rho-g}
\rho = \rho(t_{in},\mbox{\bm $x$}) 
  \frac{\sqrt{\det \bigr[ g_{ij}(t_{in},\mbox{\bm $x$}) \bigl] }}
       {\sqrt{\det \bigr[ g_{ij}(t,     \mbox{\bm $x$}) \bigl] }}\; . 
\end{equation}
Here an overdot ($\dot{\ }$) denotes $\partial/\partial t$,
and $K^i_{\ j}$ is the extrinsic curvature, whose expression
in the present gauge is $K^i_{\ j} = \frac12 g^{ik} \dot{g_{jk}}$.
Introducing the triad
\begin{equation}
g_{ij} = a^2(t) \,\delta_{(k)(\ell)} \,e^{(k)}_{\ i} e^{(\ell)}_{\ j} \;,
\end{equation}
Eq.~(\ref{rho-g}) is rewritten as
\begin{equation}
\label{rho-e}
\rho = \rho_b
  \frac{\det \bigr[ e^{(\ell)}_{\ i}(t_{in},\mbox{\bm $x$}) \bigl] }
       {\det \bigr[ e^{(\ell)}_{\ i}(t,     \mbox{\bm $x$}) \bigl] } \;.
\end{equation}

We obtain perturbative solutions for the triad $e^{(\ell)}_{\ i}$
regardless of the energy density $\rho$ up to second order
in the following form~\cite{rmkb}
\begin{equation}\label{second}
e^{(\ell)}_{\ i} = k^{(\ell)}_{\ i} + E^{(\ell)}_{\ i}
                                    + \varepsilon^{(\ell)}_{\ i} \;,
\end{equation}
where $k^{(\ell)}_{\ i}$ is the background triad defined by
$k_{ij}=\delta_{(k)(\ell)}\,k^{(k)}_{\ i} k^{(\ell)}_{\ j}$, and
$E^{(\ell)}_{\ i}$ and $\varepsilon^{(\ell)}_{\ i}$
are the first-order and the second-order solutions given by
\begin{equation}
E^{(\ell)}_{\ i} = k^{(\ell)}_{\ j}
                 \left(\frac{10}{9} \Psi \,\delta^j_{\ i}
                 + t^{\frac23}\,\Psi^{|j}_{\ |i}
                 \right) \;, \quad
\varepsilon^{(\ell)}_{\ i} = k^{(\ell)}_{\ j}
                           \left(t^{\frac23}\, \psi^j_{\ i}
                           + t^{\frac43}\, \varphi^j_{\ i}
                           \right) \;.
\end{equation}
Here $\Psi=\Psi(\mbox{\bm $x$})$ is the same function as the one used
in Eqs.~(\ref{lif}) and~(\ref{tomi}),
and $\psi^i_{\ j}=\psi^i_{\ j}(\mbox{\bm $x$})$ 
and $\varphi^i_{\ j}=\varphi^i_{\ j}(\mbox{\bm $x$})$ are
quadratic quantities of $\Psi$, written by
\begin{equation}
\psi^i_{\ j} = \frac59 \Psi^{|k} \Psi_{|k} \,\delta^i_{\ j} 
             - \frac{20}{9}      \left(\Psi\Psi^{|i}_{\ |j}
             + \Psi^{|i}\Psi_{|j}\right) \;,
\end{equation}
\begin{equation}
\varphi^i_{\ j} = \frac{3}{14}\left(
(\Psi_{\ |k}^{|k})^2-\Psi_{\ |\ell}^{|k}\Psi_{\ |k}^{|\ell}
                              \right)\delta_{\ j}^i
-\frac67 \left(
\Psi_{\ |k}^{|k}\Psi_{\ |j}^{|i}-\Psi_{\ |k}^{|i}\Psi_{\ |j}^{|k}
         \right) \;.
\end{equation}
Note that we removed a remaining gauge freedom in the linear level
to derive the above solution. (See Appendix A of Ref.~\cite{rmkb}.)
And we again neglected contributions of the decaying scalar mode
and the tensor mode.
We find that the solution~(\ref{second}) is consistent
in the metric level with Eqs.~(\ref{lif}) and (\ref{tomi})~\cite{rmkb}.

Relativistic Zel'dovich and post-Zel'dovich approximations
are obtained by substituting Eq.~(\ref{second})
into Eq.~(\ref{rho-e}).
\begin{eqnarray}\label{del-za}
\delta_{ZA} &=& \left(
\det \!\biggm[ \delta^i_{\ j} + \frac{t^{\frac23} \,\Psi^{|i}_{\ |j}}
                                     {1 + \frac{10}{9} \Psi}
       \biggm]  \right)^{-1} -1 \;, \\
\label{del-pza}
\delta_{PZA} &=& \left(
\det \!\biggm[ \delta^i_{\ j} + 
     \frac{t^{\frac23} \,\Psi^{|i}_{\ |j} + t^{\frac23} \,\psi^i_{\ j}
                                          + t^{\frac43} \,\varphi^i_{\ j}}
          {1 + \frac{10}{9} \Psi}
       \biggm]   \right)^{-1} -1 \;.
\end{eqnarray}
Abbreviations ZA and PZA denote
the Zel'dovich and the post-Zel'dovich approximations.

As was written previously, the results of the conventional linear
and second-order theories are
\begin{eqnarray}\label{del-lin}
\delta_{LIN} &=& -t^{\frac23} \Psi^{|k}_{\ |k} \;, \\
\label{del-sec}
\delta_{SEC} &=& -t^{\frac23} \Psi^{|k}_{\ |k}
                 +\frac59 t^{\frac23} \left(
                 \Psi^{|k} \Psi_{|k} + 6\Psi \Psi^{|k}_{\ |k}
                                      \right)
                 +\frac17 t^{\frac43} \left(
 5(\Psi^{|k}_{\ |k})^2 + 2\Psi^{|k}_{\ |\ell}\Psi^{|\ell}_{\ |k}
                                      \right) \;.
\end{eqnarray}
Abbreviations LIN and SEC denote
the linear and the second-order perturbation theories.
%If we expand Eqs.~(\ref{del-za}) and (\ref{del-pza}) in the Taylor series,
%we can reproduce Eqs.~(\ref{del-lin}) and (\ref{del-sec}), respectively.
Expanding Eqs.~(\ref{del-za}) and (\ref{del-pza}) under the condition
$||\Psi|| \ll 1$ (where $||\Psi||$ denotes an appropriate norm of a
function $\Psi$), Eqs.~(\ref{del-lin}) and (\ref{del-sec}) can be also 
obtained, respectively. 
In this sense, ZA and PZA are extensions of LIN and SEC
to $|\delta| \sim 1$.

%%%%%%%%%%%%%%%%%%%%%%%%%%%%%%%%%%%%%%%%%%%%%%%%%%
\section{Spherically symmetric model}\label{spher}
%%%%%%%%%%%%%%%%%%%%%%%%%%%%%%%%%%%%%%%%%%%%%%%%%%
In this section, we consider the spherically symmetric model
of gravitational instability in the FLRW universe.
There exists an exact solution known as the LTB solution,
which includes three arbitrary functions,
in the spherically symmetric case~\cite{landau}.
Here we make clear the relations between the arbitrary functions
included in the LTB solution and the ones which appears
in the approximation methods mentioned in the section~\ref{perturb}.
The line element of the LTB solution is
\begin{equation}\label{LTB_metric}
 ds^2 = -dt^2 + \frac{r'^2}{1 + f} \;dR^2 
              + r^2 (d\theta^2 + \sin^2 \theta \,d\phi^2) \;,
\end{equation}
where $(\ ') \equiv \partial/\partial R$ and
$f = f(R)$ is an arbitrary function which is related to initial velocity
of dust. $r = r(t,R)$ satisfies the following differential equation
\begin{equation}\label{dotr2}
 {\dot r}^2 = \frac{F(R)}{r} + f(R)
\end{equation}
with an arbitrary function $F(R)$, which represents initial distribution
of matter. Eq.~(\ref{dotr2}) can be integrated as follows

(i) $f > 0$:
\begin{equation}\label{f>0}
 r = \frac{F}{2f} (\cosh \eta - 1)\, , \quad
 t - t_0(R) = \frac{F}{2f^{3/2}} (\sinh \eta - \eta) \;,
\end{equation}

(ii) $f < 0$:
\begin{equation}\label{f<0}
 r = \frac{F}{-2f} (1 - \cos \eta)\, , \quad
 t - t_0(R) = \frac{F}{2(-f)^{3/2}} (\eta - \sin \eta) \;,
\end{equation}

(iii) $f = 0$:
\begin{equation}\label{f=0}
 r = \left( \frac{9F}{4} \right)^{\frac13} (t - t_0(R))^{\frac23} \;,
\end{equation}
where $t_0(R)$ is an integration constant.
The above cases (i), (ii) and (iii) may be called ``open,'' ``closed,''
and ``flat,'' respectively, as in the FLRW universe.
In these three cases, the density reads
\begin{equation}\label{rholtb}
 8\pi G\rho = \frac{F'}{r' r^2} \;.
\end{equation}
Apparently the LTB solution includes the three arbitrary functions,
$f(R)$, $F(R)$, and $t_0(R)$.
But dynamical degree of freedom is actually two
because there remains freedom of choice of the radial coordinate $R$.

The LTB solution is some extension of the FLRW solution and
is often used as a model of an inhomogeneous universe
(see Ref.~\cite{krasin} for review),
and this solution can be reduced to the FLRW solution.
By choosing $f=0$ and $t_0=0$ (and $F=\frac49 R^3$ for convenience),
we easily find that $r=a(t)R$ and the LTB solution is reduced to 
the spatially flat FLRW solution~(\ref{FLRW}).
Furthermore, we can represent this solution by the form of the flat
FLRW solution with its perturbations and can see that the
arbitrary functions in the LTB solution corresponds to those
in the linear perturbation theory.
To see this, we consider spherical linear perturbation
of the spatially flat FLRW solution. 
Substituting 
\begin{equation}
f = f_{(1)} \;, \quad t_0 = t_{0\,(1)} \;, \quad
F = \frac49 R^3 + F_{(1)} \;, \quad r = aR + r_{(1)} \;,
\end{equation}
into Eq.~(\ref{dotr2}),
the linearized equation for $r_{(1)}$ can be obtained,
where the quantities with subscript $(1)$ are treated
as linear perturbation.
Solving this equation, we obtain
\begin{equation}
r_{(1)} = aR \left( \frac34 F_{(1)} R^{-3} 
        +\frac{9}{20} t^{\frac23} f_{(1)} R^{-2} +t^{-1} B
             \right) \;,
\end{equation}
where $B=B(R)$ is an integration constant.
Actually $B$ is related to $t_{0\,(1)}$ by $B=-\frac23 t_{0\,(1)}$.
It is easily seen by choosing $f=0$ and using Eq.~(\ref{f=0}),
which reads
\begin{equation}
r_{(1)} = aR \left( \frac34 F_{(1)} R^{-3} - \frac23 t^{-1} t_{0\,(1)}
             \right) \;.
\end{equation}
Then we can write ``linearized LTB metric'' in terms of $f_{(1)}$,
$t_{0\,(1)}$, and $F_{(1)}$ as follows
\begin{eqnarray}
\gamma_{RR} &=& 1+\frac32 (F_{(1)} R^{-2})'-f_{(1)}
                 +\frac{9}{10} t^{\frac23} (f_{(1)} R^{-1})'
                 -\frac43 t^{-1} (t_{0\,(1)} R)' \;, \nonumber \\
\label{ltb1st}
\gamma_{\theta\theta} &=& R^2 \left(
       1+\frac32 F_{(1)} R^{-3}+\frac{9}{10} t^{\frac23} f_{(1)} R^{-2}
        -\frac43 t^{-1} t_{0\,(1)} \right) \;.
\end{eqnarray}
On the other hand, the solution of the linear theory~(\ref{lif}) gives
\begin{eqnarray}
\gamma_{RR} &=& 1+\frac{20}{9}\Psi +2t^{\frac23}\Psi''
                                   +2t^{-1}\Phi'' \;, \nonumber \\
\label{lif-spher}
\gamma_{\theta\theta} &=& R^2 \left(
       1+\frac{20}{9}\Psi +2t^{\frac23}\Psi' R^{-1} 
        +2t^{-1}\Phi' R^{-1}   \right) \;.
\end{eqnarray}
Comparing Eqs.~(\ref{ltb1st}) and~(\ref{lif-spher}),
we find the following relations
\begin{equation}
F_{(1)} = \frac{40}{27} \Psi R^3 \;, \quad
f_{(1)} = \frac{20}{9} \Psi' R \;, \quad
t_{0\,(1)} = -\frac32 \Phi' R^{-1} \;.
\end{equation}
This tells us that the arbitrary functions $f(R)$ and $t_0(R)$
correspond to the growing and the decaying modes, respectively,
in the linear level.
%Thus w
We choose $t_0(R) =0$ hereafter because contributions
of the decaying mode are not taken into account throughout.
Moreover, we know that $F_{(1)}$ and $f_{(1)}$ are related to each other
by the function $\Psi$.
This is because we eliminated a residual gauge freedom and thus
fixed the gauge condition completely in the previous section.
This fixing corresponds to determination of the choice
of the radial coordinate $R$ in the spherical case.

To see the relation between the LTB solution and the second-order
perturbation, we introduce $f_{(2)}$, $F_{(2)}$, and $r_{(2)}$
so that
\begin{equation}
f = \frac{20}{9} \Psi' R + f_{(2)} \;, \quad
F = \frac49 R^3 + \frac{40}{27} \Psi R^3 + F_{(2)} \;, \quad
r = aR + r_{(1)} + r_{(2)} \;,
\end{equation}
and make calculations in the same way.
Here $f_{(2)}$, $F_{(2)}$, and $r_{(2)}$ should be regarded as
$O(||\Psi||^{2})$ and $r_{(2)}$ is obtained
by solving Eq.~(\ref{dotr2}) perturbatively.
Then we obtain the part of $O(||\Psi||^{2})$ in $\gamma_{RR}$ and
$\gamma_{\theta\theta}$. 
Comparing these $\gamma_{RR}$ and $\gamma_{\theta\theta}$ 
with the solution of the second-order theory~(\ref{tomi}),
we find
\begin{equation}
F_{(2)} = \frac{400}{243} \Psi^2 R^3 \;, \quad
f_{(2)} = \frac{100}{81} (\Psi'^2 R^2 -2\Psi \Psi' R) \;.
\end{equation}
On the other hand, the LTB solution as an exact model is obtained
by choosing 
\begin{eqnarray}
F &=& \frac49 R^3 + \frac{40}{27} \Psi R^3 
                  + \frac{400}{243} \Psi^2 R^3 \;, \\
f &=& \frac{20}{9}\Psi' R + \frac{100}{81}(\Psi'^2 R^2-2\Psi \Psi' R) \;.
\end{eqnarray}
 From Eq.~(\ref{rholtb}), the density contrast of the LTB solution reads
\begin{equation}
\delta_{LTB} = \frac34 \,\frac{t^2 F'}{r' r^2} -1 \;.
\end{equation}

Let us turn our attention to the approximation methods.
If we impose $\Psi = \Psi(R)$, Eqs.~(\ref{del-lin}), (\ref{del-sec}),
(\ref{del-za}) and (\ref{del-pza}) become
\begin{eqnarray}
 \delta_{LIN} &=& -t^{\frac23} (\Psi'' + 2 \Psi' R^{-1}) \;, \\
 \delta_{SEC} &=& -t^{\frac23} (\Psi'' + 2 \Psi' R^{-1})
     + \frac59 t^{\frac23} \left(\Psi'^2 + 6\Psi (\Psi''+2\Psi' R^{-1})
                           \right)
     + t^{\frac43} \left(
 \Psi''^2 + \frac{20}{7}\Psi''\Psi'R^{-1} + \frac{24}{7}\Psi'^2 R^{-2}
                   \right) ,
\end{eqnarray}
\begin{eqnarray}
 \delta_{ZA} &=& \left(1 + \frac{t^{\frac23} \Psi'R^{-1}}
                                {1 + \frac{10}{9}\Psi} \right)^{-2}
                 \left(1 + \frac{t^{\frac23} \Psi''}
                                {1 + \frac{10}{9}\Psi} \right)^{-1}
                 - 1 \;, \hspace{40mm} \\
 && \nonumber \\
 \delta_{PZA} &=& \left(1 + \frac{t^{\frac23} \Psi'R^{-1}
       + \frac59 t^{\frac23} \left(\Psi'^2 - 4 \Psi\Psi'R^{-1}\right)
       - \frac37 t^{\frac43} \Psi'^2 R^{-2}}
                                 {1 + \frac{10}{9}\Psi} \right)^{-2}
\nonumber
\end{eqnarray}
\begin{equation}
\hspace*{15mm}
 \times \left(1 + \frac{t^{\frac23} \Psi''
       - \frac59 t^{\frac23} \left(3\Psi'^2 + 4 \Psi\Psi''\right)
       - \frac37 t^{\frac43} \Psi'R^{-1} (2 \Psi'' - \Psi'R^{-1})}
                       {1 + \frac{10}{9}\Psi} \right)^{-1}
       - 1 \;.
\end{equation}
Peculiar velocity, which represents deviation of motion
of dust shell from the Hubble expansion and is defined
by $v \equiv {\dot r}-Hr$,
(where $H \equiv {\dot a}/a$ is the Hubble parameter)
is written as follows
\begin{eqnarray}
v_{LIN} = v_{ZA}  &=& \frac23 t^{\frac13} \Psi' \;, \\
v_{SEC} = v_{PZA} &=& \frac23 t^{\frac13} \Psi'
                   + \frac{10}{27} t^{\frac13}(\Psi'^2 R - 4\Psi\Psi')
                   - \frac47 t \Psi'^2 R^{-1} \;.
\end{eqnarray}
Note that $v_{LIN} = v_{ZA}$ and $v_{SEC} = v_{PZA}$
because, in the metric level, the Zel'dovich-type approximations
coincide with the conventional ones.

Now the density contrast and the peculiar velocity
of the LTB solution and the approximations
are written in terms of the only one function $\Psi$.
The function $\Psi$ should be determined from initial conditions
so that the regularity conditions at $R=0$, i.e.,
$\Psi(R=0)=0$ and $\Psi'(R=0)=0$ are satisfied.
Then the peculiar velocity at $R=0$ is always zero
both in the LTB solution and the approximation.
Moreover, if $\Psi$ is taken so that $\Psi \propto R^2$ near $R=0$,
the peculiar velocity near $R=0$ is proportional to $R$, $v \propto R$.

%%%%%%%%%%%%%%%%%%%%%%%%%%%%%%%%%%%%%%%%%%%%%%%%%%%%%%%%
\section{Comparison of LTB solution and approximations}
\label{comparison}
%%%%%%%%%%%%%%%%%%%%%%%%%%%%%%%%%%%%%%%%%%%%%%%%%%%%%%%%

Let us proceed to comparison of the LTB solution and the approximations.
As mentioned in the section~\ref{perturb}, $\delta_{ZA}$ and $\delta_{PZA}$
include $\delta_{LIN}$ and $\delta_{SEC}$, respectively,
when the density contrast is small.
As for the peculiar velocity, ZA and PZA are coincident
with LIN and SEC, respectively.
Moreover, expanding the density contrast of ZA in the following form
\begin{equation}\label{expandza}
\delta_{ZA} \simeq -t^{\frac23} (\Psi'' + 2 \Psi'R^{-1})
      + \frac{10}{9} t^{\frac23} \Psi (\Psi'' + 2 \Psi'R^{-1})
      + t^{\frac43} (\Psi''^2 + 2\Psi''\Psi'R^{-1} + 3\Psi'^2 R^{-2})
      + O(||\Psi||^3) \;,
\end{equation}
it is found that ZA includes second-order (and higher) terms partially
in the expression of the density contrast.
Thus we can expect that $\delta_{ZA}$ is as accurate as $\delta_{SEC}$
at late time.

In order to investigate relative accuracy of the approximations
quantitatively, we compare the LTB solution and the approximations
by using some specific initial conditions.
Here initial conditions can be completely fixed
by giving an initial density profile $\delta_{in}(R)$.
For simplicity, we assume that $\delta_{in}(R)$ is a first-order quantity.
Then the arbitrary function $\Psi$ is determined by the relation
\begin{equation}\label{deltain}
\delta_{LIN} \biggm|_{t=t_{in}} = -(\Psi''+ 2\Psi' R^{-1})
                                = \delta_{in}(R)
\end{equation}
with normalization $t_{in}=1$.
($t_{in}$ may be regarded as the decoupling time in the history
of the expanding universe.)
Eq.~(\ref{deltain}) is solved to determine the function $\Psi$
with the boundary conditions $\Psi(R=0)=0$ and $\Psi'(R=0)=0$.

Here we consider the following two cases:
\begin{equation}
  \label{former_case}
\delta_{in}(R) = \epsilon \left( 1+\frac{R}{R_0} \right)
                 \exp \left( -\frac{R}{R_0} \right)
\end{equation}
and
\begin{equation}
  \label{later_case}
\delta_{in}(R) = \epsilon \left[
                 1+\frac{R}{R_0}-\left(\frac{R}{R_0}\right)^2
                          \right]
                 \exp \left( -\frac{R}{R_0} \right) \;,
\end{equation}
where $\epsilon$ is a small constant $(|\epsilon| \ll 1)$
which represents amplitude of an initial density perturbation,
and $R_0$ is a comoving scale of the fluctuations.
The former case can be regarded as smoothing out the top-hat model
while in the latter case, if $\epsilon <0$, the neighborhood of $R=0$
is underdense region (void) and the outside is overdense,
and then the shell-crossing will occur.
(If $\epsilon >0$, the latter case also shows only similar behavior
to the top-hat model as the former one.)
For both of the cases, we can evaluate $\delta$
at the center of the fluctuations $(R=0)$ from the LTB solution
and the approximation methods in the following form
\begin{equation}\label{delltbR=0}
 \delta_{LTB} (R=0) = \left\{ \begin{array}{ll}
 \frac{\displaystyle 9}{\displaystyle 2} \;\frac{\displaystyle (\eta -
   \sin \eta)^2}{\displaystyle (1 - \cos \eta)^3} - 1 \;,
 & \quad \mbox{for} \quad \epsilon > 0 \;, \\
 \frac{\displaystyle 9}{\displaystyle 2} \;\frac{\displaystyle (\eta -
   \sinh \eta)^2}{\displaystyle (\cosh \eta - 1)^3} - 1 \;, 
 & \quad \mbox{for} \quad \epsilon < 0 \;,
 \end{array} \right.
\end{equation}
with
\[ 
 t = \left\{ \begin{array}{ll}
 \frac{\displaystyle 9}{\displaystyle 20} \sqrt{\frac{\displaystyle
     3}{\displaystyle 5}} \,\epsilon^{-\frac32} \,(\eta - \sin \eta)
 \;,
 & \quad \mbox{for} \quad \epsilon > 0 \;, \\
 \frac{\displaystyle 9}{\displaystyle 20} \sqrt{\frac{\displaystyle
     3}{\displaystyle 5}} \,(-\epsilon)^{-\frac32} 
 \,(\sinh \eta - \eta) \;,
 & \quad \mbox{for} \quad \epsilon < 0 \;,
 \end{array} \right.
\]
and
\begin{eqnarray}
 \label{dellinR=0}
 \delta_{LIN} (R=0) &=& \epsilon \, t^{\frac23} \;, \\
 \label{delsecR=0}
 \delta_{SEC} (R=0) &=& \epsilon \, t^{\frac23} 
                      + \frac{17}{21} \,\epsilon^2 \,t^{\frac43} \;, \\
 \label{delzaR=0}
 \delta_{ZA} (R=0) &=& \left(\, 1 - \frac{\epsilon}{3} \,t^{\frac23}\,
                       \right)^{-3} -1 \;,\\
 \label{delpzaR=0}
 \delta_{PZA} (R=0) &=& \left(\, 1 - \frac{\epsilon}{3} \,t^{\frac23}
                                   - \frac{\epsilon^2}{21} \,t^{\frac43}\,
                        \right)^{-3} -1 \;.
\end{eqnarray}
These expressions do not depend on the details of initial density profiles.
It is of essence in the calculation that $\delta_{in}\simeq \epsilon$
near $R=0$, and $\epsilon >0$ and $\epsilon <0$ correspond to
$f<0$ and $f>0$ at $R=0$, respectively.
The above results are described in Figures 1 and 2,
where Eqs.~(\ref{delltbR=0})-(\ref{delpzaR=0}) are plotted
as functions of $\delta_{LTB}(R=0)$.
Figure 1, which represents the collapse case,
tells us that ZA is more accurate than LIN
and PZA is more accurate than SEC in all $\delta_{LTB}(R=0) >0$ region.
(Figure 1 shows only the region of $0< \delta_{LTB}(R=0) <1$,
but the tendency shown in Figure 1 does not change
in denser region $\delta_{LTB}(R=0) >1$.)
We also find that ZA becomes more accurate than SEC at late time
when $\delta_{LTB}$ is larger than about $0.5$.
This is understood by comparing Eq.~(\ref{delsecR=0}) and
the expanded form of Eq.~(\ref{delzaR=0}),
\begin{equation}
  \label{expandza_origin}
\delta_{ZA}(R=0) \simeq \epsilon \, t^{\frac23}
           + \frac23\,\epsilon^2 \, t^{\frac43} + O(\epsilon^3) \;.
\end{equation}
(This form is also obtained from Eq.~(\ref{expandza}).)
We see from Eqs.~(\ref{delsecR=0}) and (\ref{expandza_origin})
that $\delta_{ZA}$ is smaller than $\delta_{SEC}$
at early time ($\epsilon t^{2/3} \ll 1$) due to the lack of the terms
in $O(\epsilon^{2})$. In this sense, $\delta_{ZA}$ is less accurate
than $\delta_{SEC}$ when $\epsilon t^{2/3} \ll 1$.
However, due to the existence of the singularity at $\epsilon t^{2/3} = 3$
in $\delta_{ZA}$,
$\delta_{ZA}$ becomes to be more accurate than $\delta_{SEC}$ at late time.
This existence of the singularity in $\delta_{ZA}$ essentially
determines the asymptotic behavior of $\delta_{ZA}$.
Indeed, the exact solution $\delta_{LTB}$ for $\epsilon > 0$
in Eq.~(\ref{delltbR=0}) has a pole of order three at
$\epsilon t^{2/3} = 3 (3\pi/2)^{2/3}/5 \sim 1.7$ ($\eta = 2\pi$).
This pole corresponds to the crunching time at $R=0$,
and the singularity occurs at this time.

 From Eqs.~(\ref{delltbR=0})-(\ref{delpzaR=0}), we can also evaluate
accuracy of the approximations quantitatively at the turnaround time
$\eta = \pi$ ($\epsilon t^{2/3} = 3 (9\pi^{2}/2)^{1/3}/10 \sim 1.1$),
though it is not drawn in Figure 1.
Here the turnaround time is characterized by ${\dot r}=0$, i.e.,
the maximum expansion.
Physically speaking, the density fluctuation begins to collapse due to
gravitational instability, overcoming the cosmic expansion
at the turnaround time.
At the turnaround time, $\delta_{LTB}$ becomes to be $4.6$.
To this $\delta_{LTB}$, $\delta_{LIN}$, $\delta_{SEC}$,
$\delta_{ZA}$ and $\delta_{PZA}$ grow to about $23\%$, $43\%$, $60\%$
and $84\%$, respectively.
It is natural that $\delta_{PZA}$ is more accurate
than $\delta_{ZA}$ from the viewpoint of the singularity at
the crunching time. $\delta_{PZA}$ also has a pole of order $3$
at $\epsilon t^{2/3} = (\sqrt{133} -7)/2 \sim 2.3$.
This crunching time is nearer to the real crunching time $\sim 1.7$
than that of $\delta_{ZA}$.

 From Figure 2, which denotes the void case,
we find that PZA gives the best fit at early time
before $\delta_{LTB}\simeq -0.7$ while ZA works best at late time
when $\delta_{LTB}$ is smaller than about $-0.7$.
At late time, PZA gives bad results.
It is due to difference of the signature between the first-order
(-$\epsilon t^{2/3} /3 >0$) and the second-order
(-$\epsilon^2 t^{4/3} /21 <0$) terms when $\epsilon <0$.
The same feature also appears in SEC but at earlier time.
However, in PZA, this difference is more serious than in SEC. In SEC,
$\delta_{SEC}$ grows as $\epsilon^2 t^{4/3} /21 <0$, while in PZA,
the difference of the signature makes $\delta_{PZA}$ diverge
at a finite time $\epsilon t^{2/3} = (7 + \sqrt{133})/2$.
This divergence is an apparent one which is caused
by the formalization of PZA.
Indeed, we can easily see from Eq.~(\ref{delltbR=0}) that the exact
solution $\delta_{LTB}$ has no pole in $t>0$
(i.e. there is no singularity)
and approaches to $-1$ as $\sim t^{-1}$.
Though there is no singular point in $\delta_{LIN}$ except $t = \infty$,
$\delta_{LIN}$ takes value smaller and smaller 
without limit as the time increases because there is no physics to
stop this decrease of the energy density in this order.
Then, only $\delta_{ZA}$ predicts true asymptotic value of $\delta$
without growth and apparent singularities.
But $\delta_{ZA}$ approaches to $-1$ as $t^{-2}$.
This difference is also seen in Figure 2.
In the Newtonian case, detailed discussions on the void
can be seen in Ref.~\cite{sasha}.
Our relativistic results up to now are quite similar to Newtonian ones,
which are given in Refs.~\cite{bouchet,munshi,sasha}.

Figures 3~(a) and 4~(a) give the density contrast as a function of
$R/R_0$ when $\epsilon = 1.0\times 10^{-3}$ and $t=2.0\times 10^4$,
and $\epsilon = -1.0\times 10^{-3}$ and $t=3.7\times 10^4$
with the initial density profile~(\ref{former_case}), respectively.
These two figures show that difference between the LTB solution
and the approximations is the largest at $R=0$ in the former case.
Hence, it is sufficient to consider the difference at $R=0$
when we examine the accuracy of the approximations
in the former case~(\ref{former_case}).

We also see the evolution of the peculiar velocity with the initial
density profile~(\ref{former_case}) in Figures 3~(b) and 4~(b). 
Here we consider the peculiar velocity normalized by the Hubble flow $Hr$.
It will be convenient to use it to see the deviation of the model
from the FLRW universe in the metric level.
For example, the normalized peculiar velocity $v/Hr =-1$
at the turnaround time.
Figures 3~(b) and 4~(b) show the normalized peculiar velocity
corresponding to Figures 3~(a) and 4~(a).
Although these figures show that the normalized peculiar velocity
$v/Hr$ is not zero at $R=0$,
we must note that this is due to our normalization.
As mentioned in the last section, the peculiar velocity 
must behave $\sim R$ near $R=0$.
On the other hand, $Hr$ also behaves $\sim R$ near $R=0$.
Hence our normalized peculiar velocity does not
vanish at $R=0$ due to the normalization. 
It is also noted that the peculiar velocity obtained from
the Zel'dovich-type approximations is the same as the one
obtained from the conventional approximations as mentioned
in the previous section.
Furthermore, we find from Figures 3~(b) and 4~(b) that the deviation
from the FLRW universe is maximum near $R=0$
for the initial profile~(\ref{former_case}).
And this is also shows that it is sufficient to consider the
difference at $R=0$ when we examine the accuracy of the approximations
in the former case~(\ref{former_case}) as mentioned above.

On the other hand, for the initial density profile given by
Eq.~(\ref{later_case}), it cannot be said that the largest deviation of
the approximations from the LTB solution occur at $R=0$.
Figures 5~(a) and 6~(a) are for this initial profile
when $\epsilon = -1.0\times 10^{-3}$ and $t=2.0\times 10^5$,
and $\epsilon = -1.0\times 10^{-3}$ and $t=3.0\times 10^5$, respectively.
For this initial density profile, the shell-crossing singularity will occur.
The tendency of the occurring of shell-crossing can be seen
from the peculiar velocity in Figures 5~(b) and 6~(b),
where the profile of the normalized peculiar velocities for this case are drawn.
In Figures 5~(b) and 6~(b),
there exists a $v = 0$ point at $R/R_0 \sim 2.5$.
This point, which is denoted by $R = R_{c} \ne 0$ hereafter, is a boundary
where the universe is locally ``open'' ($f>0$) and ``closed'' ($f<0$),
and then $f=0$ at the point.
The peculiar velocity in the void region $R<R_{c}$ is positive,
while that in the closed region $R>R_{c}$ is negative.
Then we can expect that the shell-crossing of the dust matter will form
at $R=R_{c}$ within a finite time.

Indeed, one can see from Eqs.~(\ref{dotr2}) and~(\ref{f=0})
that the shell-crossing, which is characterized by a finite radius
at which $r'$ vanishes~\cite{Lake}, will occur at the radius $R=R_{c}$.
 From Eq.~(\ref{f=0}), which is the solution when $f=0$, we know
\begin{equation}\label{r_c}
r_c = \left(\frac{9 F_c}{4}\right)^{\frac13} t^{\frac23} \;.
\end{equation}
(Subscript $c$ denotes value at $R=R_c$.)
Differentiating Eq.~(\ref{dotr2}) with respect to $R$ and
using Eq.~(\ref{r_c}), we obtain the equation for $r'_c$.
Integrating this equation, one finds
\begin{equation}
  \label{rcdash}
  r_{c}' = \frac{3F_{c}'}{4}\left(\frac{4}{9F_{c}}\right)^{\frac23}
  t^{\frac23} + \frac{3}{5} \left(\frac{4}{9F_{c}}\right)^{\frac13}
  f_{c}' \, t^{\frac43} + C t^{-\frac13} \;,
\end{equation}
where $C$ is an integration constant and is not essential
in our argument. 
Here, $r'$ must be positive initially if $r$ is a monotonically
increasing function of $R$ which is in our case. This means that the first 
term (cooperate with the third term) in Eq.~(\ref{rcdash}) must dominate on 
the initial surface.
%Eq.~(\ref{rholtb}) shows that the energy condition $\rho>0$ on the 
%initial surface leads to $F_{c}'>0$. $r'$ must be positive when $t \ll1$.
However, since $f_{c}'<0$ at $R = R_{c}$, $r'_c$ must vanish within a 
finite time.
Hence, at $R = R_{c}$, the shell-crossing singularity will occur.
(More generic arguments about the occurrence of shell-crossing 
singularity can be seen in Ref.~\cite{Lake} and the shell-crossing may
occur in the region $f<0$ at first.)

It should be noted that, in Figures 5~(a) and 6~(a),
$\delta_{ZA}$ and $\delta_{PZA}$ take the same value
as that of the LTB solution at $R=R_{c}$
where the shell-crossing will occur.
Then we must say that these figures show that the Zel'dovich-type
approximations are not necessarily inaccurate
even when the shell-crossing is occurring. 
Let us consider the reason here. 
In fact, the deviation from the background Hubble expansion is
locally one-dimensional at the point.
The definition of ``locally one-dimensional deviation'' we adopt here is
that two of the eigenvalues of the peculiar deformation tensor
$V^i_{\ j} \equiv K^i_{\ j} - H \delta^i_{\ j} =
\frac12 \gamma^{ik} \dot{\gamma_{jk}}$ are zero~\cite{kasai93}.
According to the definition, let us show the local one-dimensionality
at $R=R_c \ne 0$. From Eqs.~(\ref{LTB_metric}) and (\ref{r_c}),
\begin{equation}
V^R_{\ R} \biggm|_{R=R_c} =
\frac{\dot r'_{\ c}}{r'_{\ c}} - \frac{\dot a}{a} \ne 0 \;, \quad
V^{\theta}_{\ \theta} \biggm|_{R=R_c} =
V^{\phi}_{\ \phi} \biggm|_{R=R_c} = 
\frac{\dot r_c}{r_c} - \frac{\dot a}{a} = 0 \;;
\quad v_{LTB} \biggm|_{R=R_c} =0 \;.
\end{equation}
Since $V^i_{\ j}$ is diagonal in the spherical case,
this means that the deviation at $R=R_c$ is locally one-dimensional.
It is known that the Zel'dovich-type approximations become exact
when the deviation is locally one-dimensional~\cite{kasai95,kasai93}.
In this argument, the origin $R=0$ must be excluded because all
components of the peculiar velocity vanish at the origin. Thus we can
lead a significant consequence that, 
at the points where $f=0$, the Zel'dovich-type approximations
coincide with the exact LTB solution, i.e.,
\begin{equation}
\delta_{LTB} = \delta_{ZA} = \delta_{PZA} \quad \mbox{and} \quad
v_{LTB} = v_{ZA} = v_{PZA} = 0 \;.
\end{equation}
Note that this consequence is not limited to the specific initial
density profile~(\ref{former_case}) nor (\ref{later_case}).

Turning to Figures 5~(a) and 6~(a), we see that the coincidence
of the density contrast at $R/R_0 \sim 2.5$ ($R = R_{c}$) contribute
to accuracy of the Zel'dovich-type approximations,
and $\delta_{ZA}$ and $\delta_{PZA}$
give good fit around $R/R_0 \sim 2.5$.
This is the reason the Zel'dovich-type approximations do not
necessarily give bad results even when the shell-crossing is occurring.

%%%%%%%%%%%%%%%%%%%%%%%%%%%%%%%%%%%%%%%%%%%%%%%%%
\section{Summary and Discussions}\label{summary}
%%%%%%%%%%%%%%%%%%%%%%%%%%%%%%%%%%%%%%%%%%%%%%%%%
We have tested the relativistic perturbative approximations 
to gravitational instability with the LTB solution.
It has been shown that the Zel'dovich-type approximations give higher
accuracy than the conventional ones in the quasi-nonlinear regime
$|\delta| \sim 1$ within general relativistic framework.
Our results are partly similar to the Newtonian ones,
but our consideration is more generic.
Especially we considered some cases in which matter distribution
is inhomogeneous, and found that the Zel'dovich-type approximations are
not necessarily inaccurate even when the shell-crossing is occurring.
Of course, the occurrence of the shell-crossing shows the break down of
our treatment. However, this is due to the failure of our description of 
the matter as dust rather than the failure of Zel'dovich-type approximations.

Indeed, one of the case considered in the previous section
includes the $f=0$ point, where the universe is
locally ``open'' inside and is locally ``closed'' outside and the 
shell-crossing will occur at this radius.
We have seen, in general, at the $f=0$ points (except the origin $R=0$),
the deviation from the FLRW model is locally one-dimensional
and the Zel'dovich-type approximations become exact.
And in the neighborhood of the points, we can expect
that the Zel'dovich-type approximations are particularly accurate.
The case considered here is exactly such an example.
It should be also noted that ``one-dimensionality'' which makes
the Zel'dovich-type approximations exact means not only globally
plane-symmetric but also locally one-dimensional:
Such situations appear even in the spherically symmetric model.

To discuss applicable range of the Zel'dovich-type approximations,
we reconsider the density contrast at $R=0$ in the collapse case.
At the turnaround time, accuracy of the Zel'dovich-type approximations
already begins to fall down, i.e., $\delta_{PZA}$ is about $84\%$
of $\delta_{LTB}$ and $\delta_{ZA}$ is about $60\%$.
Inaccuracy will be accelerated beyond the turnaround time.
In this sense, 
the turnaround epoch, when the peculiar velocity is as large as
the Hubble expansion, is one of criterion of applicable range
of the Zel'dovich-type approximations. 
However, this criterion might not be practical, because one cannot know the 
correct turnaround time in general, while we have been able to know that 
from the exact solution in our case. Instead of the turnaround time, 
the Zel'dovich-type approximations tell us the crunching time as the 
singularity in the density contrast, approximately. Furthermore, we have 
also seen that PZA tells us this crunching time more accurately than ZA 
in our spherical model. Then we may be able to know the approximate
turnaround time by the half of this crunching time of PZA.
%This is a reflection of the fact that the peculiar velocity is
%obtained in conventionally perturbative manner
%while the density contrast is computed by an extrapolation
%in the Zel'dovich-type approximations.

It is said that the Zel'dovich approximation predicts pancake formation
in the gravitational collapse of dust~\cite{zel70}.
But it is beyond the turnaround time that the pancake will be formed
and thus accuracy of the Zel'dovich approximation is not ensured
at that time.
If we try to examine final stage of the collapse quantitatively,
we will need to develop a new approximation scheme
which gives an accurate description even beyond the turnaround epoch.

\acknowledgments
M. M. would like to thank A. Hosoya for continuous encouragement,
H. Ishihara for helpful discussions, and M. Morikawa and A. Yoshisato
for valuable remarks.

%%%%%%%%%%%%%%%%%%%%%%%%%%%%%%%%%%%%%%%

%%%%%%%%%%%%%%%%%%%%%%%%%%%%%%%%%%%%%%%
\begin{figure}  %FIGURE 1
\begin{center}
\figurenum{1}
  \leavevmode
  \epsfbox{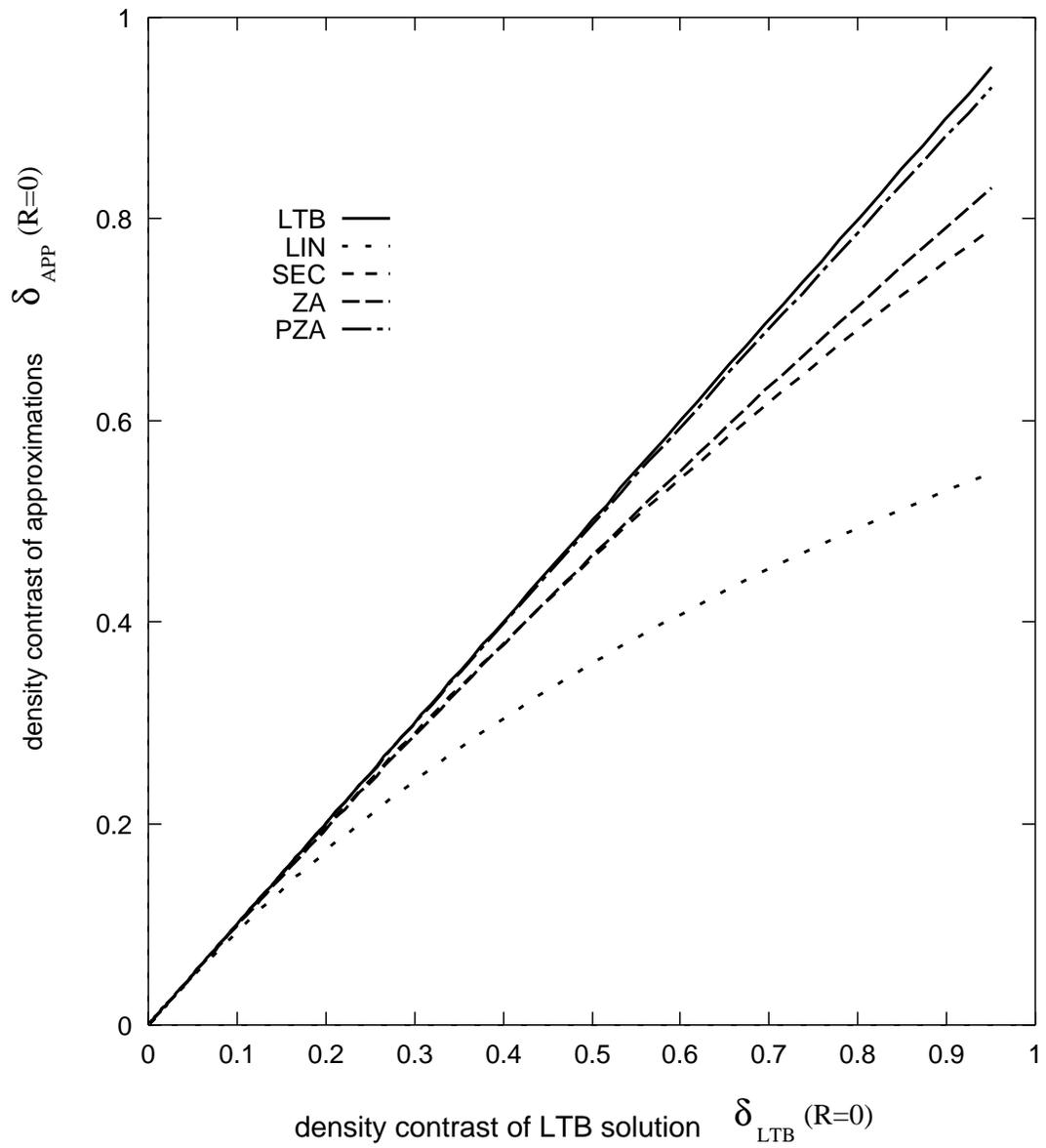}
\caption{Density contrast $\delta$ at $R=0$ calculated from
         approximation methods as a function of $\delta_{LTB}(R=0)$
         for $\epsilon >0$.
         $\delta_{ZA}(R=0)$ catches up with $\delta_{SEC}(R=0)$ at
         $\delta_{LTB}(R=0) \sim 0.5$.}
\end{center} 
\end{figure}
\begin{figure}  %FIGURE 2
\begin{center}
\figurenum{2}
  \leavevmode
  \epsfbox{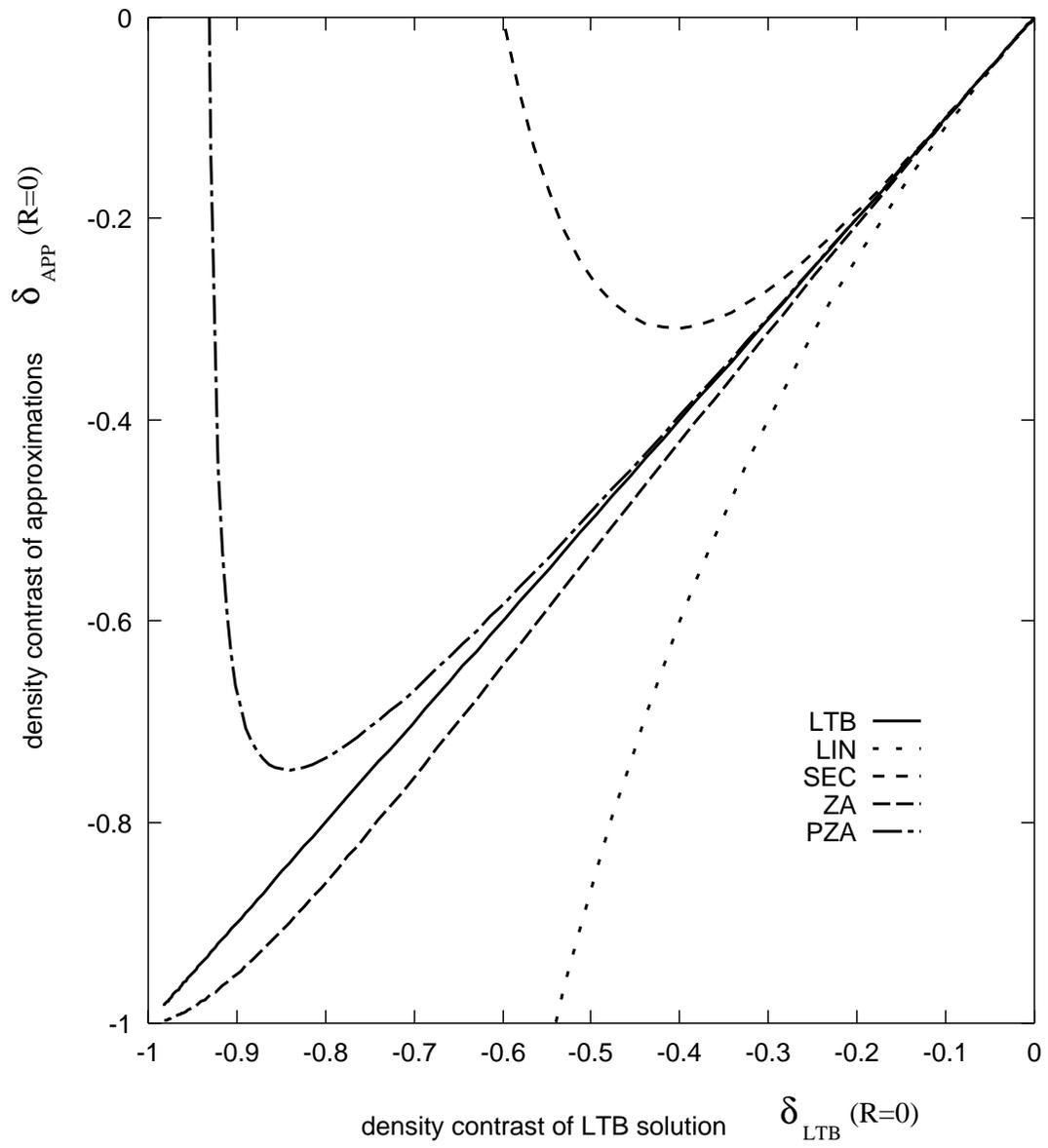}
\caption{Same as in Fig. 1, but for $\epsilon <0$.}
\end{center}
\end{figure}
\begin{figure}  %FIGURE 3 (a)
\begin{center}
\figurenum{3 (a)}
  \leavevmode
  \epsfbox{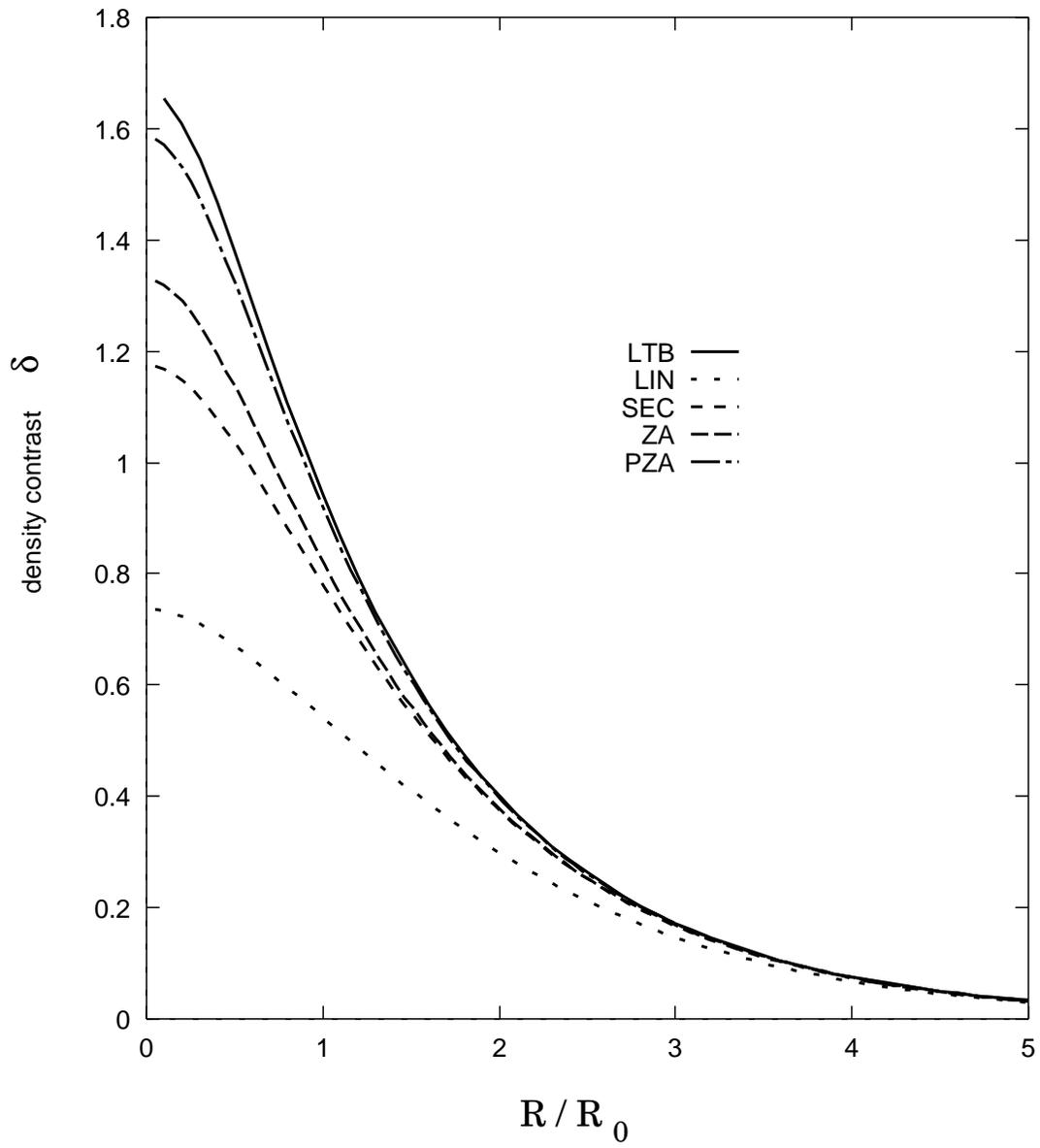}
\caption{Profile of density contrast when 
         $\delta_{in}=\epsilon(1+R/R_0)\exp(-R/R_0)$
         and $t=2.0\times 10^4$
         with $\epsilon = 1.0\times 10^{-3}$.}
\end{center}
\end{figure}
\begin{figure}  %FIGURE 3 (b)
\begin{center}
\figurenum{3 (b)}
  \leavevmode
  \epsfbox{propecc20.eps}
\caption{Profile of normalized peculiar velocity
         corresponding to Figure 3~(a).}
\end{center} 
\end{figure}
\begin{figure} %FIGURE 4 (a)
\begin{center}
\figurenum{4 (a)}
  \leavevmode
  \epsfbox{prodelv37.eps}
\caption{Same as in Figure 3~(a), but for void case.
         We choose $\epsilon =-1.0\times 10^{-3}$ and
         $t=3.7\times 10^4$.}
\end{center} 
\end{figure}
\begin{figure} %FIGURE 4 (b)
\begin{center}
\figurenum{4 (b)}
  \leavevmode
  \epsfbox{propecv37.eps}
\caption{Profile of normalized peculiar velocity
         corresponding to Figure 4~(a).}
\end{center} 
\end{figure}
\begin{figure}  %FIGURE 5 (a)
\begin{center}
\figurenum{5 (a)}
  \leavevmode
  \epsfbox{prodels200.eps}
\caption{Profile of density contrast when 
         $\delta_{in}=\epsilon(1+R/R_0-(R/R_0)^2)\exp(-R/R_0)$
         and $t=2.0\times 10^5$
         with $\epsilon = -1.0\times 10^{-3}$.}
\end{center}
\end{figure}
\begin{figure} %FIGURE 5 (b)
\begin{center}
\figurenum{5 (b)}
  \leavevmode
  \epsfbox{propecs200.eps}
\caption{Profile of normalized peculiar velocity
         corresponding to Figure 5~(a).}
\end{center}
\end{figure}
\begin{figure}  %FIGURE 6 (a)
\begin{center}
\figurenum{6 (a)}
  \leavevmode
  \epsfbox{prodels300.eps}
\caption{Same as in Figure 5~(a), but for $t=3.0\times 10^5$.}
\end{center}
\end{figure}
\begin{figure} %FIGURE 6 (b)
\begin{center}
\figurenum{6 (b)}
  \leavevmode
  \epsfbox{propecs300.eps}
\caption{Profile of normalized peculiar velocity
         corresponding to Figure 6~(a).}
\end{center} 
\end{figure}

\end{document}